\documentclass[traditabstract,rnote]{aa}
\usepackage{epsfig} 
\usepackage{graphicx}
\usepackage{amssymb}
\usepackage{natbib}
\usepackage{times}
\usepackage{txfonts}
\def\int{{\em INTEGRAL}}
\def\XMM{{\em XMM-Newton}}

\def\swift{{\em Swift}}

\def\j16479{IGR\,J16479-4514}

\def\igrj11{IGR\,J11215-5952}

\def\2s{2S\,0114+65}
 
\begin{document}

\title{{\em Swift} observations of IGR\,J16479-4514 in outburst} 

\author{E. Bozzo\inst{1,2}{\thanks{\email: bozzo@oa-roma.inaf.it}} 
\and A. Giunta\inst{2,3}  
\and L. Stella\inst{2}
\and M. Falanga\inst{4}  
\and G. Israel\inst{2} 
\and S. Campana\inst{5}}

\offprints{E. Bozzo}
\titlerunning{IGR\,J16479-4514 in outburst}
\authorrunning{E. Bozzo et al.}

\institute{ISDC, Geneva Observatory, University of Geneva, Chemin d'Ecogia 16, CH-1290 Versoix, Switzerland. 
\and  INAF - Osservatorio Astronomico di Roma, Via Frascati 33,00044 Rome, Italy.
\and Dipartimento di Fisica - Universit\`a di Roma Tor Vergata, via
della Ricerca Scientifica 1, 00133 Rome, Italy.  
\and International Space Science Institute, Hallerstrasse 6, CH-3012 Bern, Switzerland.
\and INAF – Osservatorio Astronomico di Brera, via Emilio Bianchi 46, I-23807 Merate (LC), Italy. 
}

\abstract{}{ The supergiant fast X-ray transient source \j16479\ was observed in outburst 
two times with \swift.\ Its quiescent state was investigated in-depth only once in 2008 through a 
relatively long pointed observation with \XMM.\ The latter observation was taken about 
1.7~days after the outburst in 2008, and showed an X-ray eclipse-like event, likely caused 
by the supergiant companion.  
At present, this is the only supergiant fast X-ray transient that displayed an evidence for an X-ray eclipse.}   
{Here we carry out a comparison between the most recent outburst of \j16479,\ caught by \swift\ on 29  
January 2009 and those detected previously from this source.} 
{The decay from the outbursts in 2005, 2008 and 2009 presents many 
similarities, and suggests a common mechanism that modulates the mass accretion 
rate onto the neutron star in \j16479.\ }{}       


\date{Received: 22 March 2009 / Accepted: 04 June 2009}

\maketitle

\section{Introduction}
\label{sec:intro}

\j16479\ is one of the supergiant fast X-ray transients (SFXT), a subclass of  
high mass X-ray binaries (HMXBs) hosting an accreting neutron star (NS) 
and a supergiant companion \citep{sguera06,walter06}.   
These sources undergo sporadic outbursts 
(peak luminosities of $\sim$10$^{36}$-10$^{37}$~erg~s$^{-1}$), with typical durations 
of few thousands of seconds, and remain in quiescence for most of the time, with a 
mean X-ray luminosity in the $\sim$10$^{32}$-10$^{34}$~erg~s$^{-1}$ range.  
In order to explain the behaviour of the SFXT sources, a number of different scenarios have been proposed
\citep{zand05,walter07,negueruela08,sidoli07,bozzo08}. All these models  
involve a NS accreting from the intense wind of its supergiant companion, and interpret the   
transitions from outburst to quiescence as being due to strong variations in the mass accretion rate 
onto the NS. These variations have been ascribed to 
a centrifugal and/or magnetic gating mechanism by the NS magnetosphere \citep{bozzo08}, 
or to clumps and structures in the wind of the supergiant star \citep{walter07, negueruela08}. 
\citet{sidoli07} proposed that SFXT outbursts result from a 
neutron star in a tilted orbit crossing a dense equatorial disk around the supergiant companion. 
As different models make different predictions with respect to the origin of the quiescent emission in  
SFXTs, an in-depth knowledge of the timing and spectral properties of the low activity states of these sources 
is expected to help discriminating between the above models \citep{bozzo08}. 
At present, only a few observations of quiescent SFXTs have been carried out, and thus very little is 
known on the mechanism that triggers the transitions between outburst and quiescence. 

\j16479\ was observed with \swift\ in outburst two times, on 2005 August 30 \citep{sguera08} and 2008 March 
19 \citep{romano08,romano08b}. Its X-ray emission properties during these bright events is fairly well known.  
On the contrary, the quiescent state of \j16479\ was investigated in-depth only once with \XMM\ \footnote{\citet{walter06} 
also reported on a $\sim$2.9~ks \XMM\ observation of this source in quiescence, but it was affected by a very strong and variable 
background.}.   
This observation lasted $\sim$32~ks and was carried out about 1.7~days after the bright outburst in 2008, with the aim  
of gaining insight on the mechanism that drive the quiescent emission of this source.  
Surprisingly, \XMM\ caught \j16479\ undergoing a transition from  
a bright state (2-10~keV absorbed flux 1.0$\times$10$^{-11}$~erg/cm$^2$/s) to a lower flux state 
(2-10~kev absorbed flux 7.5$\times$10$^{-13}$~erg/cm$^2$/s). The spectral and timing analysis 
of this observation showed unambiguously that the transition was due to the obscuration 
by an occulting body, most likely an eclipse by the supergiant companion. 
This revealed that at least some of the very large X-ray luminosity swings 
displayed by the SFXT sources might not be due only to genuine changes in the mass 
accretion rate. 

{Here we report on the most recent outburst of \j16479,\ detected by \swift\ on 2009 January 29,      
and compare this outburst with those observed previously in 2005, 2008. }

\section{Data analysis and results}
\label{sec:observation}

\swift\ /BAT caught a new bright outburst from \j16479\ on 2009 
January 29 \citep{romano09}. The start time of the BAT trigger was 06:33:07 UTC  
(including UTCF correction). \swift\ /XRT slewed to the source about 800~s 
after the BAT trigger, and observed in Windowed Timing (WT) mode during the 
first 46~s. Data in photon counting (PC) mode were then 
accumulated for a total exposure time of 2.6~ks. 
We analyzed the data from observation ID. 00341452000, that includes the BAT 
data of the outburst, as well as the following XRT WT and PC data, and use  
also \swift\ follow-up observations that were carried out up to 20~days after the 
BAT trigger. 

In order to compare the 2009 outburst with those detected previously, 
we also analyzed \swift\ observations of this source that were 
carried out during the outbursts in 2005 and 2008 
\citep[these observations were previously published by][]{sguera08,romano08}. 
We used outburst and follow-up observations up to $\sim$20~days  
after each BAT trigger. 
A log of the data sets is given in Table~\ref{tab:observations}. 

The \swift\ data were analyzed by using standard procedures 
\citep{burrows05} and the latest calibration files available. 
The BAT and XRT data were processed with the {\sc batproduct} (v.2.42)   
and {\sc xrtpipeline} (v.0.12.1) tasks, respectively.  
Filtering and screening criteria were applied 
by using {\sc ftools} (Heasoft v.6.6.1). 
We extracted source and background light 
curves and spectra by selecting event grades of 0-2 and 0-12,  
respectively for the WT and PC mode. Exposure maps were created through 
the {\sc xrtexpomap} task, and we used the latest spectral redistribution 
matrices in the {\sc heasarc} calibration database (v.011). 
Ancillary response files, accounting for different 
extraction regions, vignetting and PSF corrections, were generated 
by using  the {\sc xrtmkarf} task.  
When required, we corrected PC data for pile-up, and used the 
{\sc xrtlccorr} to account for this correction in the 
background-subtracted light curves. 
\begin{figure}
\centering
\includegraphics[scale=0.34,angle=-90]{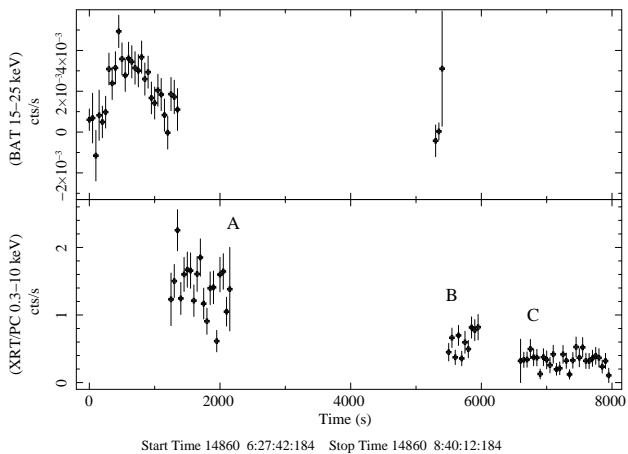}
\caption{BAT and XRT/PC light curve of the 2009 outburst (time bin is 50~s).}  
\label{fig:BAT} 
\end{figure}

For each observation in Table~\ref{tab:observations}, we extracted the light curve and spectrum, 
and derived a mean X-ray flux by fitting this spectrum with an absorbed power law model 
(we used {\sc Xspec} v.12.5.0). 
Spectra with an adequate statistics were rebinned in order to 
have at least 20 photons per bin so as to permit $\chi^2$ fitting. 
Lower statistic spectra were rebinned in order to have at least 5 photons 
per bin and then fit by using the C-statistics \citep{cash79}.    
Observations with less than 100~counts were treated separately, as the C-statistics 
is not recommended with such a small number of 
photons\footnote{https://astrophysics.gsfc.nasa.gov/XSPECwiki/statistical\_methods \_in\_XSPEC.}. 
In these cases we estimated the source count rate of the observation with {\sc sosta} ({\sc ximage} V.4.4.1), 
and then used this count rate within {\sc webpimms}\footnote{http://heasarc.nasa.gov/Tools/w3pimms.html}  
in order to derive the X-ray flux (we assumed the same spectral model 
of the closest observation for which a spectral analysis could be carried out).  

\begin{figure}
\centering
\includegraphics[scale=0.35,angle=-90]{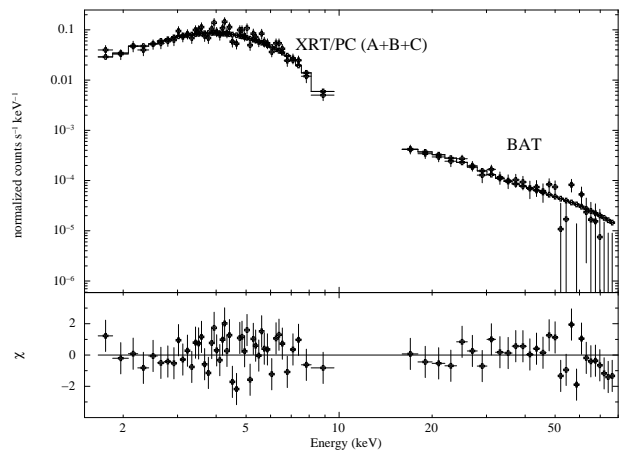}
\caption{BAT and XRT/PC spectra of the 2009 outburst. The BAT spectrum is integrated over 
the total outburst duration, whereas the XRT/PC spectrum is integrated over the 
A+B+C time intervals shown in Fig.~\ref{fig:BAT}. The former spectrum was fit 
by using a simple power law, whereas for the latter we also included the effect of the  
absorption at low energies (the best fit parameters are reported in 
Table~\ref{tab:observations}). The lower panel shows the residuals from the best fits.}  
\label{fig:total_spectrum} 
\end{figure}

Figure~\ref{fig:BAT} shows the BAT (15-25~keV) and XRT/PC (0.3-10~keV) light curve of the 2009 outburst  
(XRT/WT data in the 0.3-10~keV energy band are included only in Fig.~\ref{fig:lcurve}). 
As the observation interval of BAT and XRT/PC do not significantly overlap, we did not attempt to fit a combined  
BAT+XRT spectrum. Instead we fit them separately and report the results of these fits 
in Table~\ref{tab:observations}. The BAT spectrum, integrated over the total duration of the outburst,  
and the XRT/PC spectrum of observation ID 00341452000 (A+B+C intervals) are shown in Fig.~\ref{fig:total_spectrum}, 
together with the best fits and the residuals from these fits. We also searched for spectral variability 
by extracting time resolved spectra in the intervals A, B, and C of Fig.~\ref{fig:BAT}. 
This analysis revealed that the source X-ray flux decreased from 3.2$\times$10$^{-10}$~erg/cm$^{2}$/s$^{-1}$ in 
interval A, to 3.6$\times$10$^{-11}$~erg/cm$^{2}$/s$^{-1}$ in interval C.  
The best fit powerlaw index and the absorption column density did not show any significant 
variation (they were all constant within the errors). 

\begin{figure}
\centering
\includegraphics[scale=0.45,angle=-90]{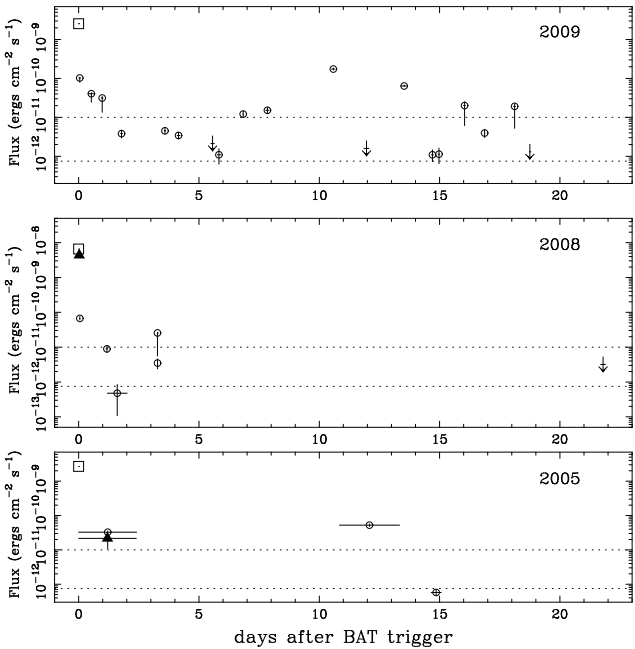}
\caption{Evolution of the X-ray flux during the 2005, 2008, and 2009 outbursts. 
We used all the observations reported in Table~\ref{tab:observations}, and considered 
the BAT trigger time as the origin of the X axis. The BAT trigger times are    
2005-08-30 04:08:48 UTC, 2008-03-19 22:44:45 UTC, and 2009-01-29 06:33:07 UTC.  
In all panels, squares indicates the BAT X-ray flux (15-80~keV for the 2009 and 2008 outburst and 
15-50~keV for the 2005 outburst; we did not indicated the errors for these fluxes), 
open circles and triangles are for XRT/PC and XRT/WT fluxes (absorbed flux in the 2-10~keV band), respectively. 
The arrows indicate 3$\sigma$ upper limits. Errors on the X axis correspond to the half 
duration of each observation.} 
\label{fig:longterm} 
\end{figure}

\begin{figure}
\centering
\includegraphics[scale=0.47]{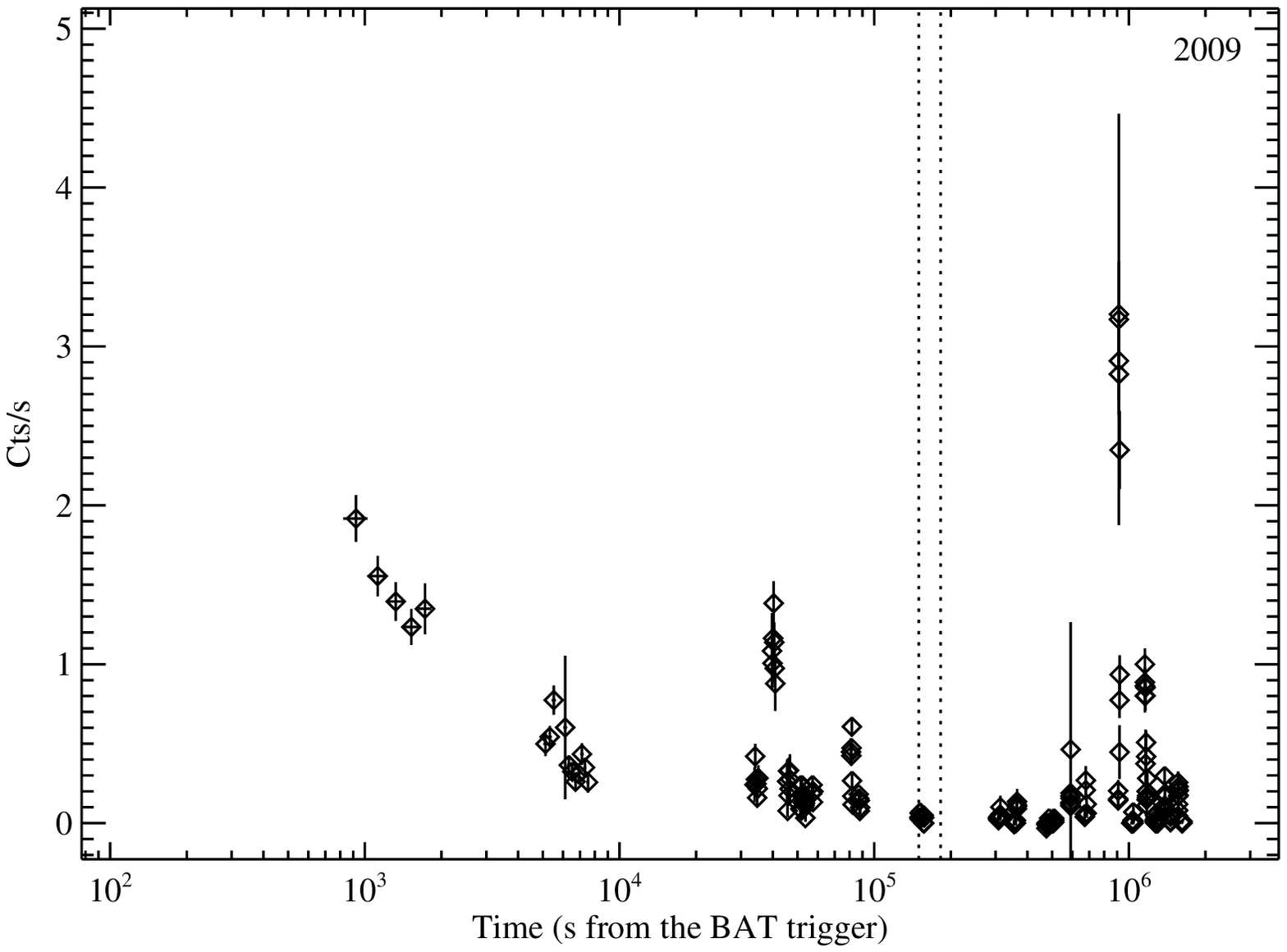}
\includegraphics[scale=0.47]{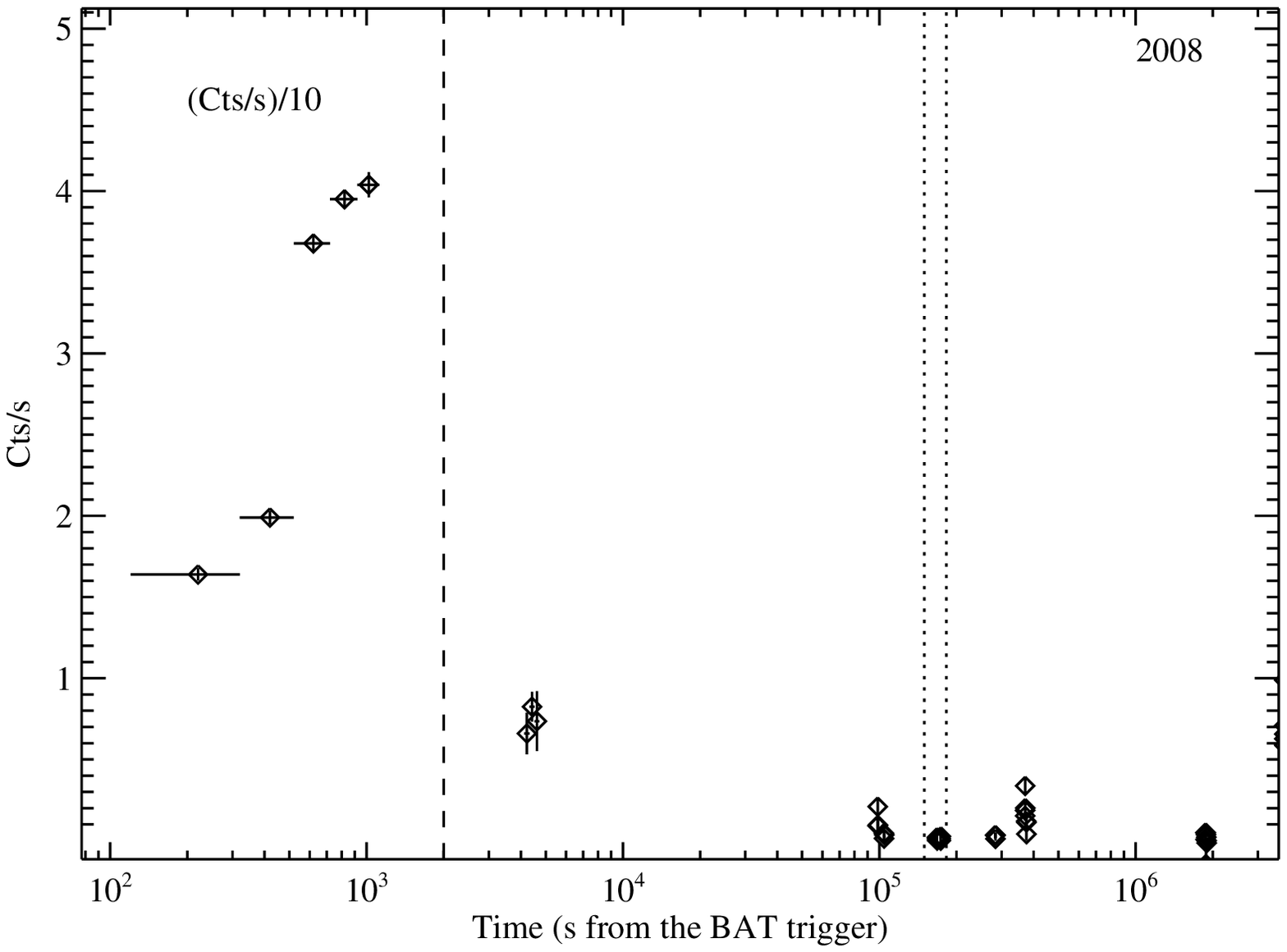}
\includegraphics[scale=0.47]{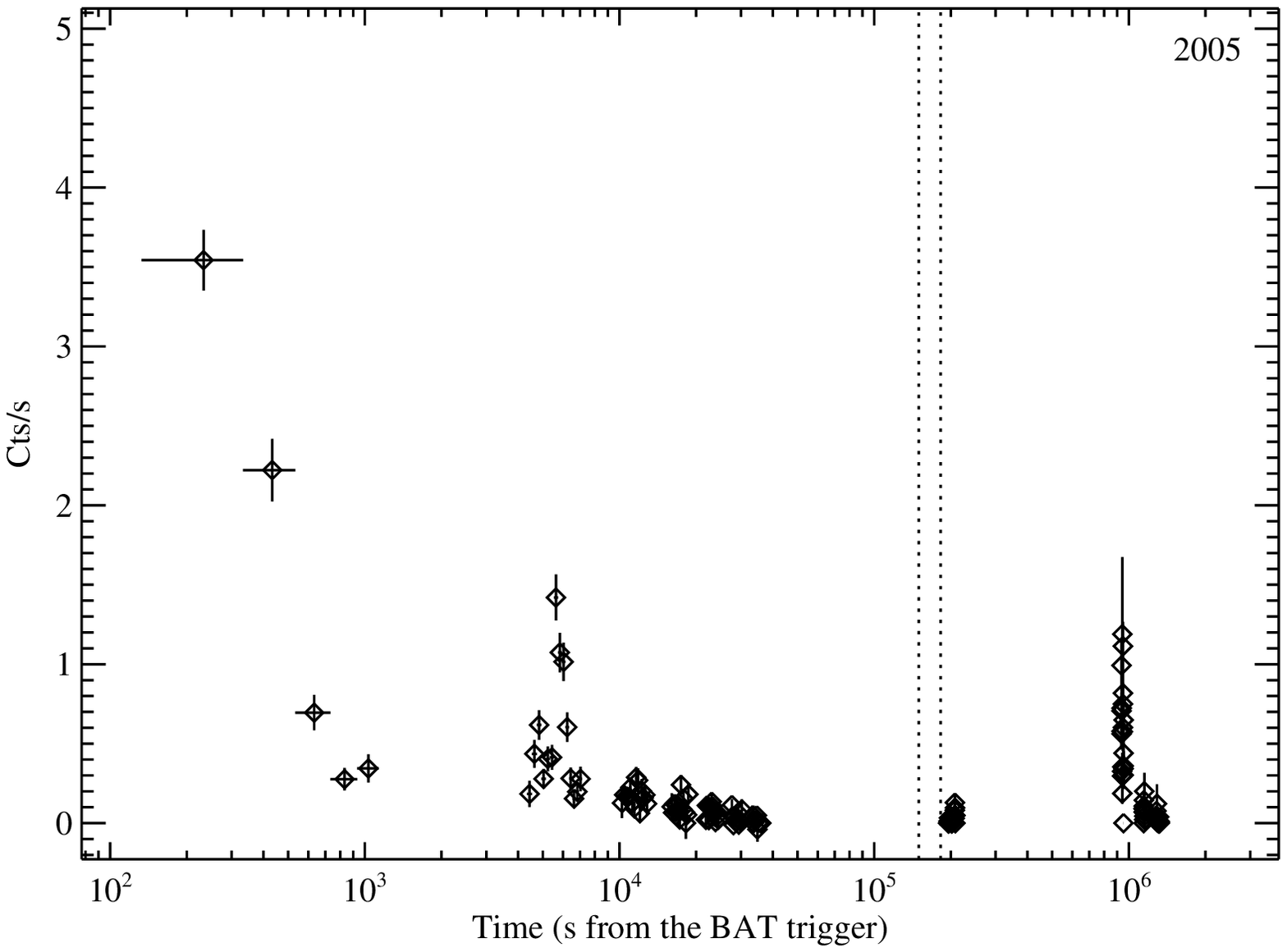}
\caption{XRT/PC and XRT/WT light curves of the three outbursts in 2005, 2008, and 2009 
(time bin is 200~s). All the reported times are measured from the BAT triggers (see the caption  
of Fig.~\ref{fig:longterm}). In the light curve of the 2008 outburst, the beginning and the end 
of the \XMM\ observation that could have indicated an X-ray eclipse are marked with dotted vertical lines. 
In the light curves of the other two outbursts, the dotted lines indicate where the suspected eclipse 
should have been observed. For clarity, we divided the count rate of the first 5 points (XRT/WT data) of 
the 2008 light curve by a factor of 10 (these points are separated from the others through a dashed line).} 
\label{fig:lcurve} 
\end{figure}

In Fig.~\ref{fig:longterm} we plot the evolution 
(up to $\sim$20~days from the BAT trigger) of the absorbed 2-10~keV X-ray flux
during the 2005, 2008 and 2009 outburst of \j16479.\ 
In this figure we used the fluxes from Table~\ref{tab:observations} and considered 
half of each observation duration as the error on the time axis. 

Finally, in Fig.~\ref{fig:lcurve} we show the combined XRT/PC and XRT/WT light curves  
of the three outburts in detail. For clarity, we divided the count rate of the first 5 
points (XRT/WT data) of the 2008 light curve by a factor of 10. 

\begin{table*}
\tiny
\caption{Observations log of IGRJ16479-4514 (see note at the bottom of the Table for other specifications).} 
\begin{tabular}{@{}lllllllll@{}}
\hline
\hline
\noalign{\smallskip}
OBS ID & INSTR & START TIME & STOP TIME & EXP & $N_{\rm H}$ & $\Gamma$ & $F_{\rm obs}$ & $\chi^2_{\rm red}$/d.o.f. \\
 (2009)    &       &            &           &  (sec)   &  (10$^{22}$~cm$^{-2}$) &  & (erg/cm$^{2}$/s$^{-1}$) & (C-stat/d.o.f.) \\
\noalign{\smallskip}
\hline
\noalign{\smallskip}
00341452000 & BAT    & 2009-01-29 06:26:12 & 2009-01-29 07:56:43 & 1.6E+03 &          ---        & 2.5$^{+0.3}_{-0.3}$ & 2.55$\times$10$^{-9}$                & 0.9/27 \\

            & XRT/WT & 2009-01-29 06:46:53 & 2009-01-29 08:16:49 & ---     &          ---        & ---                & ---                                  & --- \\

            & XRT/PC & 2009-01-29 06:47:35 & 2009-01-29 08:39:04 & 2.6E+03 & 4.7$^{+1.3}_{-1.1}$  & 1.0$^{+0.3}_{-0.3}$ & 1.0$^{+0.1}_{-0.2}\times$10$^{-10}$ & 1.1/44 \\

00030296077 & XRT/PC & 2009-01-29 15:59:44 & 2009-01-29 22:36:24 & 5.9E+03 & 5.7$^{+1.7}_{-1.5}$  & 1.6$^{+0.4}_{-0.4}$ & 4.1$^{+0.2}_{-1.6}\times$10$^{-11}$   & 1.2/33 \\

00030296078 & XRT/PC & 2009-01-30 05:05:45 & 2009-01-30 07:01:56 & 2.5E+03 & 9.3$^{+3.6}_{-3.2}$  & 1.5$^{+0.7}_{-0.6}$ & 3.1$^{+0.3}_{-1.8}\times$10$^{-11}$   & 1.2/21 \\

00030296079 & XRT/PC & 2009-01-31 00:23:45 & 2009-01-31 02:17:58 & 1.4E+03 & 9.3 (fixed)         & 1.5 (fixed)        & 3.8$^{+0.8}_{-0.8}\times$10$^{-12}$   & --- \\

00030296080 & XRT/PC & 2009-02-01 19:50:46 & 2009-02-01 21:30:57 & 2.0E+03 & 9.3 (fixed)         & 1.5 (fixed)        & 4.5$^{+0.7}_{-0.7}\times$10$^{-12}$   & --- \\

00030296081 & XRT/PC & 2009-02-02 08:46:12 & 2009-02-02 11:40:57 & 1.9E+03 & 9.3 (fixed)         & 1.5 (fixed)        & 3.4$^{+0.7}_{-0.7}\times$10$^{-12}$   & --- \\

00030296082 & XRT/PC & 2009-02-03 18:25:48 & 2009-02-03 21:49:58 & 1.2E+03 & 9.3 (fixed)$^a$        & 1.5 (fixed)$^a$       & 2.1$\times$10$^{-12}$                & --- \\

00030296083 & XRT/PC & 2009-02-04 00:51:50 & 2009-02-04 04:09:56 & 1.4E+03 & 9.3 (fixed)         & 1.5 (fixed)        & 1.1$^{+0.5}_{-0.5}\times$10$^{-12}$   & --- \\

00030296084 & XRT/PC & 2009-02-05 02:24:25 & 2009-02-05 02:50:30 & 1.5E+03 & 9.3 (fixed)         & 1.5 (fixed)        & 1.2$^{+0.2}_{-0.2}\times$10$^{-11}$   & --- \\

00030296085 & XRT/PC & 2009-02-06 01:13:12 & 2009-02-06 04:28:57 & 4.7E+02 & 9.3 (fixed)         & 1.5 (fixed)        & 1.5$^{+0.3}_{-0.3}\times$10$^{-11}$   & --- \\

00030296087 & XRT/PC & 2009-02-08 18:54:08 & 2009-02-08 22:15:57 & 1.6E+03 & 6.9$^{+2.7}_{-2.4}$  & 1.1$^{+0.6}_{-0.5}$ & 1.7$^{+0.1}_{-0.1}\times$10$^{-10}$   & 0.9/26 \\

00030296088 & XRT/PC & 2009-02-10 03:10:18 & 2009-02-10 08:06:57 & 1.5E+03 & 6.9 (fixed)$^a$        & 1.2 (fixed)$^a$       & 1.6$\times$10$^{-12}$                & --- \\

00030296089 & XRT/PC & 2009-02-11 15:47:39 & 2009-02-11 22:38:57 & 2.2E+03 & 10.6$^{+2.8}_{-2.4}$ & 2.1$^{+0.6}_{-0.5}$ & 6.4$^{+0.4}_{-0.2}\times$10$^{-11}$   & 0.9/27 \\

00030296090 & XRT/PC & 2009-02-12 22:30:18 & 2009-02-13 00:20:56 & 1.7E+03 & 10.6 (fixed)        & 2.1 (fixed)        & 1.1$^{+0.4}_{-0.4}\times$10$^{-12}$   & --- \\

00030296091 & XRT/PC & 2009-02-13 05:04:18 & 2009-02-13 06:45:54 & 1.3E+03 & 10.6 (fixed)        & 2.1 (fixed)        & 1.1$^{+0.5}_{-0.5}\times$10$^{-12}$   & --- \\

00030296092 & XRT/PC & 2009-02-14 06:34:01 & 2009-02-14 08:18:56 & 1.3E+03 & 8.3$^{+6.8}_{-5.2}$  & 1.0$^{+1.2}_{-1.1}$ & 2.0$^{+0.5}_{-1.4}\times$10$^{-11}$   & (26.46/24)$^b$ \\

00030296093 & XRT/PC & 2009-02-15 03:19:41 & 2009-02-15 03:39:57 & 1.2E+03 & 8.3 (fixed)         & 1.0 (fixed)        & 4.0$^{+0.9}_{-0.9}\times$10$^{-12}$   & --- \\

00030296094 & XRT/PC & 2009-02-16 08:26:29 & 2009-02-16 10:14:42 & 1.1E+03 & 11.1$^{+6.0}_{-4.3}$ & 2.1$^{+1.1}_{-1.0}$ & 1.9$^{+0.3}_{-1.4}\times$10$^{-11}$   & (23.41/29)$^b$ \\

00030296095 & XRT/PC & 2009-02-17 00:22:55 & 2009-02-17 00:38:50 & 9.5E+02 & 11.1 (fixed)$^a$       & 2.1 (fixed)$^a$       & 1.3$\times$10$^{-12}$                & --- \\
\noalign{\smallskip}
\hline
\hline
\noalign{\smallskip}
OBS ID & INSTR & START TIME & STOP TIME & EXP & $N_{\rm H}$  & $\Gamma$ & $F_{\rm obs}$  & $\chi^2_{\rm red}$/d.o.f. \\
 (2008)    &       &            &           &  (sec)   &  (10$^{22}$~cm$^{-2}$) &  & (erg/cm$^{2}$/s$^{-1}$) & (C-stat/d.o.f.)\\
\noalign{\smallskip}
\hline
\noalign{\smallskip}

00306829000 & BAT    & 2008-03-19 22:42:48 & 2008-03-19 22:56:50 & 6.3E+02 & ---                 & 2.8$^{+0.1}_{-0.1}$    & 6.6$\times$10$^{-9}$                & 1.5/27 \\

            & XRT/WT & 2008-03-19 22:46:47 & 2008-03-19 23:55:26 & 9.1E+02 & 6.0$^{+0.3}_{-0.3}$  & 1.1$^{+0.1}_{-0.1}$ & 4.4$^{+0.1}_{-0.1}\times$10$^{-9}$ & 1.1/284 \\

            & XRT/PC & 2008-03-19 23:55:27 & 2008-03-20 00:00:53 & 3.2E+02 & 6.0 (fixed)         & 1.1 (fixed)          & 6.7$^{+0.9}_{-0.7}\times$10$^{-11}$  & --- \\

00030296030 & XRT/PC & 2008-03-21 02:05:29 & 2008-03-21 03:48:58 & 8.1E+02 & 6.0 (fixed)         & 1.1 (fixed)          & 8.9$^{+1.9}_{-1.9}\times$10$^{-12}$ & --- \\

00030296031 & XRT/PC & 2008-03-21 21:22:28 & 2008-03-21 23:05:57 & 8.2E+02 & 6.0 (fixed)         & 1.1 (fixed)          & 4.7$^{+3.7}_{-3.7}\times$10$^{-13}$ & --- \\

00030296032 & XRT/PC & 2008-03-23 05:25:03 & 2008-03-23 05:35:30 & 6.0E+02 & 6.0 (fixed)         & 1.1 (fixed)          & 3.5$^{+1.1}_{-1.1}\times$10$^{-12}$ & --- \\

00030296033 & XRT/PC & 2008-03-24 05:36:51 & 2008-03-24 06:50:56 & 1.0E+03 & 12.2$^{+8.8}_{-7.5}$  & 1.7$^{+1.3}_{-1.2}$ & 2.6$^{+0.1}_{-2.0}\times$10$^{-11}$ & (30.9/37)$^b$ \\

00030296034 & XRT/PC & 2008-04-10 15:17:37 & 2008-04-10 20:10:57 & 8.8E+02 & 12.2 (fixed)$^a$  & 1.7 (fixed)$^a$ & 3.2$\times$10$^{-12}$ & --- \\

\noalign{\smallskip}
\hline
\hline
\noalign{\smallskip}
OBS ID & INSTR & START TIME & STOP TIME & EXP & $N_{\rm H}$ & $\Gamma$ & $F_{\rm obs}$  & $\chi^2_{\rm red}$/d.o.f. \\
 (2005)    &       &            &           &  (sec)   &  (10$^{22}$~cm$^{-2}$) &  & (erg/cm$^{2}$/s$^{-1}$) & (C-stat/d.o.f.) \\
\noalign{\smallskip}
\hline
\noalign{\smallskip}

00152652000 & BAT    & 2005-08-30 04:03:49 & 2005-08-30 04:13:51 & 6.3E+02 & ---                 & 2.0$^{+0.7}_{-0.7}$ & 2.7$\times$10$^{-9}$$^c$               & 1.2/14 \\           

00030296001 & XRT/PC & 2005-08-30 04:12:41 & 2005-09-01 14:01:58 & 1.1E+04 & 6.3$^{+1.4}_{-1.2}$  & 1.0$^{+0.3}_{-0.3}$ & 3.3$^{+0.3}_{-2.2}\times$10$^{-11}$ & 0.9/44 \\

            & XRT/WT & 2005-08-30 04:11:00 & 2005-09-01 13:40:51 & 4.5E+03 & 6.0$^{+3.0}_{-2.2}$  & 1.2$^{+0.6}_{-0.5}$ & 2.2$^{+0.3}_{-0.2}\times$10$^{-11}$ & 0.9/33 \\

00030296002 & XRT/PC & 2005-09-10 00:15:00 & 2005-09-12 12:10:59 & 6.3E+03 & 9.7$^{+3.6}_{-2.8}$  & 1.2$^{+0.6}_{-0.5}$ & 5.2$^{+0.2}_{-3.0}\times$10$^{-11}$ & 1.9/27 \\

00030296003 & XRT/PC & 2005-09-14 00:43:34 & 2005-09-14 10:47:57 & 4.2E+03 & 9.7 (fixed)         & 1.2 (fixed)        & 5.7$^{+1.0}_{-1.0}\times$10$^{-13}$ & --- \\

\noalign{\smallskip}
\hline
\hline
\noalign{\smallskip}
\multicolumn{9}{l}{NOTE: Spectra extracted from these observations are fit with an absorbed power law 
(absorption column density N$_{\rm H}$ and photon index $\Gamma$). $F_{\rm obs}$ is the} \\
\multicolumn{9}{l}{XRT/PC or XRT/WT (BAT) absorbed flux in the 0.3-10~keV (15-80~keV) energy band. 
EXP indicates the total exposure time of each observation} \\ 
\multicolumn{9}{l}{( \swift\ observations comprise several snapshots and are not continuous pointings at the source).} \\
\multicolumn{9}{l}{$^{a}$ 3$\sigma$ upper limit.}\\
\multicolumn{9}{l}{$^{b}$ C-statistic has been used to fit the spectrum.}\\
\multicolumn{9}{l}{$^{c}$ In this case the BAT flux is in the energy range 15-50~keV.}\\
\end{tabular}
\label{tab:observations}
\end{table*} 

It is seen from Fig.~\ref{fig:BAT} that the rise and decay times of the 2009  
outburst are of order of a few thousand seconds, 
similar to the time scales measured in the 2005 and 2008 outbursts \citep{sguera08,romano08}. 
The fluxes and spectral properties of the source emission during the peak 
and the beginning of the outburst decay (up to $\sim$7~ks after the BAT trigger) 
are also in agreement with previously reported results. 
The typical outburst spectrum at high energy (15-50~keV) can be fit 
by a power law of photon index $\sim$2.5. The mean X-ray flux 
is of few times 10$^{-9}$~erg/cm$^{2}$/s$^{-1}$ \citep{sguera08}.   
The 2-10 keV spectrum is flatter and can be well fit by using a power law 
of photon index $\Gamma$$\sim$1 and an absorption column density of 
$\sim$5$\times$10$^{22}$~cm$^{-2}$ \citep{romano08}. 
We did not attempt to fit the broad-band spectrum of the 2009 outburst, since the 
observation intervals of BAT and XRT do not overlap significantly (see Fig.~\ref{fig:BAT}). 
Moreover, owing to poor statistics of the XRT/PC data, time resolved spectra could not be investigated 
during the first 7~ks after the BAT trigger (intervals A, B and C in Fig.~\ref{fig:BAT}). 
Therefore, we could not search for spectral variations during the decay from the outburst. 
However, we note that follow up observations carried out up to 20~days after the BAT trigger 
showed a moderate increase of the absorption column density (a factor of $\sim$2) and a steepening  
of the power law index (from $\sim$1 to $\sim$2, see Table~\ref{tab:observations}).  
 
In order to compare the 2009 outburst with those detected previously in 2005 and 2008, we studied the 
variation of the X-ray flux from \j16479\ during the 20~days that followed each of these outbursts. This is 
shown in Fig.~\ref{fig:longterm}. 
In the middle panel of this figure we give the X-ray flux of the 2008 outburst; the horizontal 
dotted lines represent the high and lower flux state of \j16479\ as observed in 2008 by \XMM\ (see Sect.~\ref{sec:intro}). 
According to the results discussed in \citet{bozzo08b}, these correspond to  
the average X-ray flux during the eclipse ingress and the mid-eclipse, respectively. 
For comparison, these two dotted lines are also plotted in the upper and lower panels of 
Fig.~\ref{fig:longterm}, that give the evolution of the X-ray flux  
during the 2009 and 2005 outbursts, respectively.  
We note that, in all the three cases, several observations have an X-ray flux that falls inside 
the range defined by the two dotted lines. 
As a consequence of this and the relatively sparse time coverage, the plots in Fig.~\ref{fig:longterm} 
do not permit to single out the presence of an X-ray eclipse. Rather \j16479\ may simply  
alternate between higher and lower flux states; unfortunately, the origin of these states 
can hardly be investigated without a detailed spectral analysis\footnote{The 2008 \XMM\ spectra, on the contrary, 
provided convincing evidence that the lower intensity state resulted from an occultation episode.}. 

By setting the start times of all light curves in Fig.~\ref{fig:longterm} at the onset of the outbursts, we can compare 
the flux evolution and, in particular, verify whether the eclipse-like event $\sim$1.7 d after the 2008 event, 
has a counterpart in the 2005 and 2009 light curves for comparable delays. Stated differently, this amounts 
to checking whether the outburst and the eclipse-like event are driven by the same, possibly orbital, "clock".

\swift\ observations in 2008 clearly show that the source X-ray flux between 1-2~days after the BAT trigger 
was compatible with the lower flux measured by \XMM\ (i.e. the mid-eclipse state). 
Furthermore, the mean X-ray flux of the subsequent observation in 2008 (that is compatible with the eclipse 
egress, if the eclipse is symmetric), suggests an upper limit to the  
duration the eclipse-like event of about $\sim$3~days. Even thought the orbital period of \j16479\ is still 
not known, this limit seems reasonable, as in other SFXTs an orbital period of $\sim$30~days  
has been measured \citep{bird09,zurita09}, and, of course, an eclipse can last up to half 
of the orbital period.  
However, if we supposed that the eclipse-like event takes place at a fixed delay from the 
beginning of the outburst, then the 2009 observation would indicate that the eclipse duration is shorter.  
In fact, from the upper panel of Fig.~\ref{fig:longterm}  we note that the source was observed 
at an X-ray flux comparable with the eclipse egress about $\sim$2~days after the BAT trigger, thus limiting 
the eclipse duration to less than one day. A comparison with the 2005 outburst is more complicated because 
the XRT observation carried out after the BAT trigger has a large gap close to the expected 
eclipse ingress and egress (see below). 

In order to further study the source behavior during the 2005, 2008, and 2009 outbursts, we also investigated 
in detail the X-ray light curves of all observations in Table~\ref{tab:observations}. 
This is shown in Fig.~\ref{fig:lcurve} (times on the X axis are measured from the start time of the BAT 
trigger). In the middle panel of this figure we represented with dotted vertical 
lines the time interval in which the 2008 eclipse-like event was seen.  
Under the hypothesis that the X-ray eclipse of \j16479\ always follows the outburst, the two dotted lines in 
the upper and lower panel of Fig.~\ref{fig:lcurve} correspond to the position where an eclipse should have been 
observed after the outbursts in 2009 and 2005, respectively.  
Even if the poor statistics of the \swift\ data did not permit a detailed spectral analysis, 
and thus did not provide further support in favour of this hypothesis,  
we note that the source count-rate at these positions 
is consistent within all the three outbursts. 
    
Apart from the X-ray eclipse-like event, we also note from Fig.~\ref{fig:longterm} that the three outbursts 
displayed many other similarities. Particularly interesting is the fact the source showed in all the 
reported outbursts an increase in the X-ray activity after about 6~ks from the BAT trigger 
(note that, close to this time interval the higher source count rate triggered XRT observations in WT 
mode in all three cases). 
Furthermore, we note the presence of a bright flare (XRT/PC count rate $>$1~cts/s) about 
$\sim$10~days after the BAT triggers in 2009 and 2005 \citep[see also,][]{parola09}. 
The presence of this flare could not be investigated during the 
outburst in 2008, because no X-ray observation of \j16479\ was carried out at this period.

\section{Discussion and Conclusion}
\label{sec:discussions} 

The similarities we pointed out in the post-outburst decay of \j16479\ in 2005, 2008 and 2009  suggest the 
presence of a stable structure in the system (at least for a few years), 
that is able to modulate the mass accretion rate onto the NS in a fairly reproducible fashion 
and gives rise to the observed X-ray activity. 
At present, it is not clear if other SFXT display a similar post-outburst behaviour.  
If this was the case, the implications for SFXT models would have to be assessed.   

After this paper was submitted for publication, \citet{jain09} 
reported the discovery of 0.6~d long X-ray eclipse and a 3.3194~d 
orbital period in the Swift/BAT data of IGRJ16479-4514.
These results, besides confirming the interpretation by 
\citet{bozzo08} that the sudden drop-off in the source X-ray flux detected 
by \XMM\ in 2008 was due to an X-ray eclipse, provided also
a clear confirmation of the conjecture proposed in the present
research note, namely that the source outbursts of 
2005, 2008 and 2009 preceded the X-ray eclipse (in 2008) and the lowest 
X-ray fluxes intervals (in 2005 and 2009) by the same amount of time.
Indeed by using the orbital period and eclipse duration of \citet{jain09}  
and the eclipse onset time from the \XMM\ observation of 
2008 plus 0.3~d to derive the eclipse center time 
(leading to a superior conjunction epoch of MJD 54547,05418, which we adopt as orbital 
phase 0),  we verified that the onset of the 2005, 2008, and 2009 outbursts, as measured by the 
\swift\ /BAT trigger time, occurred at an orbital phase of 
0.36$\pm$0.09, 0.3654$\pm$0.0002, and 0.36$\pm$0.03 respectively. 
Moreover the 0.6~d eclipse duration determined by \citet{jain09}
agrees well with what we suggested in Sect.~\ref{sec:observation} based on the 
\swift\ /XRT data of the 2009 outburst.

These findings provide strong evidence that the source outbursts, 
though rare, take place around an orbital phase of $\sim$0.4 
and are thus inherently connected to the orbital motion, though 
the conditions that lead to their occurrence are met only rarely.  
Outbursts recurring at the orbital period were observed in three  
other SFXTs \citep[SAX\,J1818.6-1703, IGR\,J11215-5952, 
and IGR\,J18483-0311;][]{bird09, zurita09, sidoli07, sguera07}, and are 
reminiscent of Type I outbursts in Be star X-ray 
transients. These episodes of intense accretion are likely related
to periastron passage, where the supergiants' wind density is higher
and its velocity  relative to the neutron star lower \citep{stella86}. 
Alternatively the outbursts might be due to the neutron star crossing an 
equatorial dense disk ejected by the supergiant \citep{sidoli07}.

\section*{Acknowledgments}
We thank the anonymous referee for his/her helpful comments, 
and the Swift staff, for having carried out ToO and follow-up observations of \j16479.\ 
EB thanks M. Perri, M. Capalbi, and K. Page for their support during the 
{\em Swift}/XRT and {\em Swift}/BAT data analysis. 
This work was partially supported through ASI and MIUR grants.

\end{document}